# Divergence-Free Reconstruction of Magnetic Fields and WENO Schemes for Magnetohydrodynamics

By

Dinshaw S. Balsara

(dbalsara@nd.edu)

Department of Physics, University of Notre Dame,

Notre Dame, IN 46556

## Abstract

Balsara (2001, J. Comput. Phys., 174, 614) showed the importance of divergence-free reconstruction in adaptive mesh refinement problems for magnetohydrodynamics (MHD) and the importance of the same for designing robust second order schemes for MHD was shown in Balsara (2004, ApJS, 151, 149). Second order accurate divergence-free schemes for MHD have shown themselves to be very useful in several areas of science and engineering. However, certain computational MHD problems would be much benefited if the schemes had third and higher orders of accuracy. In this paper we show that the reconstruction of divergence-free vector fields can be carried out with better than second order accuracy. As a result, we design divergence-free weighted essentially non-oscillatory (WENO) schemes for MHD that have order of accuracy better than second. A multistage Runge-Kutta time integration is used to ensure that the temporal accuracy matches the spatial accuracy. Accuracy analysis is carried out and it is shown that the schemes meet their design accuracy for smooth problems. Stringent tests are also presented showing that the schemes perform well on those tests.

# 1) Introduction

The Magnetohydrodynamic (MHD) equations play an important role in many areas of astrophysics, space physics and engineering. Typical applications in those areas require one to capture flow on a range of scales in a way that is as dissipation-free as possible. As a result, there has been considerable interest in bringing accurate and reliable numerical methods to bear on this problem. The MHD system of equations can be written as a set of hyperbolic conservation laws. As a result, early efforts concentrated on straightforwardly applying second order total variation diminishing (TVD) techniques to the MHD equations. This was done by Brio & Wu [13], Zachary, Malagoli & Colella [34], Powell [27], Dai & Woodward [16], Ryu & Jones [29], Roe & Balsara [28], Balsara [1] and [2], Falle, Komissarov & Joarder [22] and Crockett et al [15]. Recent efforts have focused on understanding the structure of the induction equation:

$$\frac{\partial \mathbf{B}}{\partial t} + \nabla \times (c\,\mathbf{E}) = 0 \tag{1}$$

and the divergence-free evolution that it implies for the magnetic field. In eqn. (1), $\mathbf{B}$ is the magnetic field, $\mathbf{E}$ is the electric field and c is the speed of light. The magnetic field starts out divergence-free because of the absence of magnetic monopoles and eqn. (1) ensures that it remains divergence-free for all time. The electric field is given by:

$$c\,\mathbf{E} = -\,\mathbf{v} \times \mathbf{B} \tag{2}$$

where $\mathbf{v}$ is the fluid velocity. For the rest of this paper we will simplify the notation by making the transcription $c\,\mathbf{E} \rightarrow \mathbf{E}$. Brackbill & Barnes [11] have shown that violating the $\nabla \cdot \mathbf{B} = 0$ constraint leads to unphysical plasma transport orthogonal to the magnetic field. This comes about because violating the constraint results in the addition of extra source terms in the momentum and energy equations. Yee [33] was the first to formulate divergence-free schemes for electromagnetism. Brecht et al [12] and DeVore [19] did the same for flux corrected transport (FCT)-based MHD. Dai & Woodward [17], Ryu et al [30], Balsara & Spicer [9] and [10], Balsara [4] and Londrillo and DelZanna [26] showed that simple extensions of higher order Godunov schemes permit one to formulate divergence-free time-update strategies for the magnetic field. Balsara & Kim [6] intercompared divergence-cleaning and divergence-free schemes for numerical MHD. They found that if the test problems are made stringent enough the schemes that are based on divergence-cleaning show significant inadequacies when used for astrophysical applications. Thus it is advantageous to design robust schemes for numerical MHD that are divergence-free, as was done in Balsara [4]. Balsara [4] used the divergence-free reconstruction of vector fields from Balsara (2001) to present a formulation that overcame several inconsistencies in previous formulations. Balsara et al [5] a new class of higher order schemes for the Euler equations. In such formulations the lower moments of the solution are retained while the higher moments are reconstructed, resulting in low storage schemes with better than second order accuracy.

Higher order schemes for MHD have been attempted. Jiang & Wu [26] and Balsara & Shu [8] experimented with weighted essentially non-oscillatory (WENO) schemes. Another line of effort stems from the work of Londrillo and DelZanna [26]. These schemes were based on a finite difference formulation. For certain types of applications, especially those involving non-uniform meshes or adaptive solution strategies, finite volume formulations become essential. We therefore present a finite volume, divergence-free scheme for MHD that goes beyond second order of accuracy. We rely on efficient WENO interpolation strategies that were designed in Balsara et al [7] to make a high order reconstruction. The novel element introduced in this paper consists of extending the divergence-free reconstruction of magnetic fields from Balsara [3] and [4] to all orders up to fourth. When coupled with an appropriately accurate Runge-Kutta (RK) time integration scheme by Shu & Osher [31] and [32], we get a set of WENO schemes that have a spatial and temporal accuracy that exceeds that of second order schemes.

In Section 2 we catalogue the divergence-free reconstruction of vector fields for higher order schemes. In Section 3 we provide a step by step description of the scheme.

In Section 4 we provide an accuracy analysis and in Section 5 we present several test problems.

## 2) Higher Order Divergence-Free Reconstruction of Vector Fields

In this section we study the divergence-free reconstruction of a divergenceless vector field for schemes with better than second order accuracy. In particular, we focus on the third and fourth order cases because they can be catalogued succinctly and are likely to be generally useful. The second order accurate divergence-free reconstruction of vector fields was studied for Cartesian meshes in Balsara [3]. In Balsara [4] we extended this to logically rectangular meshes with diagonal metrics. Balsara [4] also considered the second order accurate divergence-free reconstruction of vector fields on tetrahedral meshes and that too can be extended to higher orders. Since the method was described in detail in Balsara [3], in this paper we will focus on cataloguing results for the higher order case. The reader who wants a pedagogical introduction is referred to Balsara [3] and [4].

For the rest of this work we assume that each zone has been mapped to a unit cube with local coordinates $(x,y,z) \in [-1/2, 1/2] \times [-1/2, 1/2] \times [-1/2, 1/2]$. A natural set of modal basis functions within that zone or on its faces would consist of tensor products of the Legendre polynomials $P_0(x)$, $P_1(x)$ and $P_2(x)$. The first few Legendre polynomials are given by:

$$P_0(x) = 1 \; ; \; P_1(x) = x \; ; \; P_2(x) = x^2 - \frac{1}{12} \; ; \; P_3(x) = x^3 - \frac{3}{20} x \; ;$$
$$P_4(x) = x^4 - \frac{3}{14} x^2 + \frac{3}{560}$$
(3)

The above Legendre polynomials have just been suitably scaled to the local coordinates of the zone being considered. The x-component of the magnetic field in the upper and lower x-faces of this zone can be projected into these bases as:

$$B_x(x = \pm 1/2, y, z) = B_0^{x\pm} + B_y^{x\pm} P_1(y) + B_z^{x\pm} P_1(z) \qquad \leftarrow \text{second order}$$
$$+ B_{yy}^{x\pm} P_2(y) + B_{yz}^{x\pm} P_1(y) P_1(z) + B_{zz}^{x\pm} P_2(z) \qquad \leftarrow \text{third order}$$
$$+ B_{yyy}^{x\pm} P_3(y) + B_{yyz}^{x\pm} P_2(y) P_1(z) + B_{yzz}^{x\pm} P_1(y) P_2(z) + B_{zzz}^{x\pm} P_3(z) \qquad \leftarrow \text{fourth order}$$

(4)

Here $B_0^{x\pm}$, $B_y^{x\pm}$ and $B_z^{x\pm}$ are the moments that would be needed in a second order accurate representation in the basis functions that we have chosen. $B_{yy}^{x\pm}$, $B_{zz}^{x\pm}$ and $B_{yz}^{x\pm}$ are the additional moments for a third order accurate representation in the same set of basis functions. $B_{yyy}^{x\pm}$, $B_{yyz}^{x\pm}$, $B_{yzz}^{x\pm}$ and $B_{zzz}^{x\pm}$ are the further moments that are needed for a fourth order accurate representation, again in the same set of basis functions. Consequently, while eqn. (4) shows all the facial moments that are needed up to fourth order, the arrows in eqn. (4) show the terms that are needed for each specific order of accuracy. We can write similar expressions for the y and z-components of the field in the appropriate zone faces as:

$$B_y(x, y = \pm 1/2, z) = B_0^{y\pm} + B_x^{y\pm} P_1(x) + B_z^{y\pm} P_1(z) \qquad \leftarrow \text{second order}$$
$$+ B_{xx}^{y\pm} P_2(x) + B_{xz}^{y\pm} P_1(x) P_1(z) + B_{zz}^{y\pm} P_2(z) \qquad \leftarrow \text{third order}$$
$$+ B_{xxx}^{y\pm} P_3(x) + B_{xxz}^{y\pm} P_2(x) P_1(z) + B_{xzz}^{y\pm} P_1(x) P_2(z) + B_{zzz}^{y\pm} P_3(z) \qquad \leftarrow \text{fourth order}$$

(5)

$$B_z(x, y, z = \pm 1/2) = B_0^{z\pm} + B_x^{z\pm} P_1(x) + B_y^{z\pm} P_1(z) \qquad \leftarrow \text{second order}$$
$$+ B_{xx}^{z\pm} P_2(x) + B_{xy}^{z\pm} P_1(x) P_1(y) + B_{yy}^{z\pm} P_2(y) \qquad \leftarrow \text{third order}$$
$$+ B_{xxx}^{z\pm} P_3(x) + B_{xxy}^{z\pm} P_2(x) P_1(y) + B_{xyy}^{z\pm} P_1(x) P_2(y) + B_{yyy}^{z\pm} P_3(y) \qquad \leftarrow \text{fourth order}$$

(6)

To reconstruct the field in the interior of the zone we pick the following functional forms for the fields:

$$B_x(x, y, z) = a_0 + a_x P_1(x) + a_y P_1(y) + a_z P_1(z)$$
$$+ a_{xx} P_2(x) + a_{xy} P_1(x) P_1(y) + a_{xz} P_1(x) P_1(z) \quad \leftarrow \text{second order}$$
$$+ a_{yy} P_2(y) + a_{xyy} P_1(x) P_2(y) + a_{zz} P_2(z) + a_{xzz} P_1(x) P_2(z) + a_{yz} P_1(y) P_1(z) + a_{xyz} P_1(x) P_1(y) P_1(z)$$
$$+ a_{xxx} P_3(x) + a_{xxy} P_2(x) P_1(y) + a_{xxz} P_2(x) P_1(z) \quad \leftarrow \text{third order}$$
$$+ a_{yyy} P_3(y) + a_{xyyy} P_1(x) P_3(y) + a_{yyz} P_2(y) P_1(z) + a_{xyyz} P_1(x) P_2(y) P_1(z)$$
$$+ a_{yzz} P_1(y) P_2(z) + a_{xyzz} P_1(x) P_1(y) P_2(z) + a_{zzz} P_3(z) + a_{xzzz} P_1(x) P_3(z)$$
$$+ a_{xxxx} P_4(x) + a_{xxxy} P_3(x) P_1(y) + a_{xxxz} P_3(x) P_1(z)$$
$$+ a_{xxyy} P_2(x) P_2(y) + a_{xxzz} P_2(x) P_2(z) \quad \leftarrow \text{fourth order}$$

(7)

$$B_y(x, y, z) = b_0 + b_x P_1(x) + b_y P_1(y) + b_z P_1(z)$$
$$+ b_{yy} P_2(y) + b_{xy} P_1(x) P_1(y) + b_{yz} P_1(y) P_1(z) \quad \leftarrow \text{second order}$$
$$+ b_{xx} P_2(x) + b_{xxy} P_2(x) P_1(y) + b_{zz} P_2(z) + b_{yzz} P_1(y) P_2(z) + b_{xz} P_1(x) P_1(z) + b_{xyz} P_1(x) P_1(y) P_1(z)$$
$$+ b_{yyy} P_3(y) + b_{xyy} P_1(x) P_2(y) + b_{yyz} P_2(y) P_1(z) \quad \leftarrow \text{third order}$$
$$+ b_{xxx} P_3(x) + b_{xxxy} P_3(x) P_1(y) + b_{xxz} P_2(x) P_1(z) + b_{xxyz} P_2(x) P_1(y) P_1(z)$$
$$+ b_{xzz} P_1(x) P_2(z) + b_{xyzz} P_1(x) P_1(y) P_2(z) + b_{zzz} P_3(z) + b_{yzzz} P_1(y) P_3(z)$$
$$+ b_{yyyy} P_4(y) + b_{xyyy} P_1(x) P_3(y) + b_{yyyz} P_3(y) P_1(z)$$
$$+ b_{xxyy} P_2(x) P_2(y) + b_{yyzz} P_2(y) P_2(z) \quad \leftarrow \text{fourth order}$$

(8)

$$B_z(x, y, z) = c_0 + c_x P_1(x) + c_y P_1(y) + c_z P_1(z)$$
$$+ c_{zz} P_2(z) + c_{xz} P_1(x) P_1(z) + c_{yz} P_1(y) P_1(z) \quad \leftarrow \text{second order}$$
$$+ c_{xx} P_2(x) + c_{xxz} P_2(x) P_1(z) + c_{yy} P_2(y) + c_{yyz} P_2(y) P_1(z) + c_{xy} P_1(x) P_1(y) + c_{xyz} P_1(x) P_1(y) P_1(z)$$
$$+ c_{zzz} P_3(z) + c_{xzz} P_1(x) P_2(z) + c_{yzz} P_1(y) P_2(z) \quad \leftarrow \text{third order}$$
$$+ c_{xxx} P_3(x) + c_{xxxz} P_3(x) P_1(z) + c_{xxy} P_2(x) P_1(y) + c_{xxyz} P_2(x) P_1(y) P_1(z)$$
$$+ c_{xyy} P_1(x) P_2(y) + c_{xyyz} P_1(x) P_2(y) P_1(z) + c_{yyy} P_3(y) + c_{yyyz} P_3(y) P_1(z)$$
$$+ c_{zzzz} P_4(z) + c_{xzzz} P_1(x) P_3(z) + c_{yzzz} P_1(y) P_3(z)$$
$$+ c_{xxzz} P_2(x) P_2(z) + c_{yyzz} P_2(y) P_2(z) \quad \leftarrow \text{fourth order}$$

(9)

The rationale for picking this set of moments follows from Balsara [3]. Relative to the format followed in Balsara [3], a slight rearrangement of the functional forms has been made in the previous three equations to cast them in terms of the basis functions. Analogous to eqn. (4), eqn. (7) shows the terms that have to be included to achieve second, third and fourth order accuracy. Eqns. (8) and (9) have a structure that is similar to eqn. (7) and the corresponding terms that are needed with increasing accuracy are easily identified. The procedure for enforcing the divergence-free constraint is entirely similar to the one in Balsara [3] and will not be repeated here.

We now provide the formulae for obtaining the coefficients in eqn. (7) using the coefficients in eqns. (4), (5) and (6). To obtain the coefficients in eqn. (8) make the cyclic rotation of variables, a → b, b → c, c → a, x → y, y → z and z → x, in the formulae below. Similarly, to obtain the coefficients in eqn. (9) make the cyclic rotation of variables, a → c, b → a, c → b, x → z, y → x and z → y. Note that the formulae in this Section should be implemented in code in the same sequence as described here.

The description of the fourth order divergence-free reconstruction starts with this paragraph. Matching the modal basis functions with cubic terms at the $x = \pm 1/2$ boundaries gives:

$$a_{yyy} = \frac{1}{2}\left(B_{yyy}^{x+} + B_{yyy}^{x-}\right) \quad ; a_{xyyy} = B_{yyy}^{x+} - B_{yyy}^{x-} \quad ;$$

$$a_{yyz} = \frac{1}{2}\left(B_{yyz}^{x+} + B_{yyz}^{x-}\right) \quad ; a_{xyyz} = B_{yyz}^{x+} - B_{yyz}^{x-} \quad ;$$

$$a_{yzz} = \frac{1}{2}\left(B_{yzz}^{x+} + B_{yzz}^{x-}\right) \quad ; a_{xyzz} = B_{yzz}^{x+} - B_{yzz}^{x-} \quad ;$$

$$a_{zzz} = \frac{1}{2}\left(B_{zzz}^{x+} + B_{zzz}^{x-}\right) \quad ; a_{xzzz} = B_{zzz}^{x+} - B_{zzz}^{x-} \quad ;$$

(10)

Eqn. (10) gives us the coefficients $a_{yyy}$, $a_{xyyy}$, $a_{yyz}$, $a_{xyyz}$, $a_{yzz}$, $a_{xyzz}$, $a_{zzz}$ and $a_{xzzz}$ in eqn. (7). Making the analogous match of the cubic terms at the $y = \pm 1/2$ boundaries in eqn. (8) give us $b_{zzz}$, $b_{yzzz}$, $b_{xzz}$, $b_{xyzz}$, $b_{xxz}$, $b_{xxyz}$, $b_{xxx}$ and $b_{xxxy}$. It is worth pointing out that making a cyclic rotation of the variables in eqn. (10) also yields the same coefficients that are needed in eqn. (8). Matching the cubic terms at the $z = \pm 1/2$ boundaries for eqn. (9)

gives us $c_{xxx}, c_{xxxz}, c_{xxy}, c_{xxyz}, c_{xyy}, c_{xyyz}, c_{yyy}$ and $c_{yyyz}$. Notice too that making another cyclic rotation of variables also yields the coefficients for eqn. (9). We now apply the divergence-free constraint to the quartic terms in eqns. (7) to (9). After making an SVD minimization of the integral of the reconstructed magnetic energy over the zone w.r.t. the coefficients $a_{xxxy}$, $a_{xxxz}$, $a_{xxyy}$ and $a_{xxzz}$, the resulting constraints are:

$$a_{xxxx} = -\frac{1}{4}(b_{xxxy} + c_{xxxz}) \quad ; a_{xxxy} = -\frac{7}{30} c_{xxyz} \quad ; a_{xxxz} = -\frac{7}{30} b_{xxyz} \quad ;$$
$$a_{xxyy} = -\frac{3}{20} c_{xyyz} \quad ; a_{xxzz} = -\frac{3}{20} b_{xyzz} \quad \quad \quad \quad \quad \quad \quad \quad \quad (11)$$

Notice that the right hand sides of eqn. (11) are available by this point in the computation so that eqn. (11) can be used to obtain the coefficients $a_{xxxx}$, $a_{xxxy}$, $a_{xxxz}$, $a_{xxyy}$ and $a_{xxzz}$ in eqn. (7). A cyclic rotation of variables gives us the constraints for the coefficients in eqns. (8) and yields $b_{yyyy}$, $b_{yyyz}$, $b_{xyyy}$, $b_{yyzz}$ and $b_{xxyy}$. Likewise a cyclic rotation of variables gives us the coefficients in eqn. (9) and yields $c_{zzzz}$, $c_{xzzz}$, $c_{yzzz}$, $c_{xxzz}$ and $c_{yyzz}$. All the terms that are evaluated in this paragraph will be needed in the subsequent formulae when fourth order reconstruction is carried out. However, for reconstruction at third and second orders they can all be set to zero.

The description of the third order divergence-free reconstruction starts with this paragraph. This paragraph also continues our description of the fourth order reconstruction. Matching the modal basis functions with quadratic terms at the $x = \pm 1/2$ boundaries gives:

$$a_{yy} = \frac{1}{2}(B_{yy}^{x+} + B_{yy}^{x-}) - \frac{1}{6} a_{xxyy} \quad ; a_{xyy} = B_{yy}^{x+} - B_{yy}^{x-} \quad ;$$
$$a_{yz} = \frac{1}{2}(B_{yz}^{x+} + B_{yz}^{x-}) \quad ; a_{xyz} = B_{yz}^{x+} - B_{yz}^{x-} \quad ; \quad \quad \quad \quad \quad (12)$$
$$a_{zz} = \frac{1}{2}(B_{zz}^{x+} + B_{zz}^{x-}) - \frac{1}{6} a_{xxzz} \quad ; a_{xzz} = B_{zz}^{x+} - B_{zz}^{x-}$$

Eqn. (12) provides $a_{yy}$, $a_{xyy}$, $a_{yz}$, $a_{xyz}$, $a_{zz}$ and $a_{xzz}$ all of which are needed in eqn. (7). Making a cyclic rotation of variables in eqn. (12) yields the analogous terms in eqns. (8), i.e. $b_{zz}$, $b_{yzz}$, $b_{xz}$, $b_{xyz}$, $b_{xx}$ and $b_{xxy}$, all of which are needed in eqn. (8). Likewise, another cyclic rotation of variables gives the coefficients $c_{xx}$, $c_{xxz}$, $c_{xy}$, $c_{xyz}$, $c_{yy}$ and $c_{yyz}$ that are needed in eqn. (9). We are now ready to apply the constraints on the cubic terms in eqns. (7) to (9). After making an SVD minimization of the integral of the reconstructed magnetic energy over the zone w.r.t. the coefficients $a_{xxy}$ and $a_{xxz}$ we get:

$$a_{xxx} = -\frac{1}{3}(b_{xxy} + c_{xxz}) \quad ; a_{xxy} = -c_{xyz}/4 \quad ; a_{xxz} = -b_{xyz}/4 \tag{13}$$

Eqn. (13) gives us the coefficients $a_{xxx}$, $a_{xxy}$ and $a_{xxz}$ in eqn. (7). Analogous terms in eqns. (8) and (9) can now be made via a cyclic rotation of variables so that we obtain $b_{yyy}$, $b_{yyz}$, $b_{xyy}$, $c_{zzz}$, $c_{xzz}$ and $c_{yzz}$. This paragraph again gives us all the terms that will be needed in the subsequent formulae when third or fourth order reconstruction is carried out. However, for second order divergence-free reconstruction the coefficients that have been obtained in this and the previous paragraph are set to zero.

Our description of the second order divergence-free reconstruction starts with this paragraph. The present paragraph also continues our description of the third or fourth order reconstruction. Matching the modal basis functions with linear terms at the $x = \pm 1/2$ boundaries gives:

$$a_y = \frac{1}{2}\left(B_y^{x+} + B_y^{x-}\right) - \frac{1}{6} a_{xxy} \quad ; a_{xy} = \left(B_y^{x+} - B_y^{x-}\right) - \frac{1}{10} a_{xxxy} \quad ;$$
$$a_z = \frac{1}{2}\left(B_z^{x+} + B_z^{x-}\right) - \frac{1}{6} a_{xxz} \quad ; a_{xz} = \left(B_z^{x+} - B_z^{x-}\right) - \frac{1}{10} a_{xxxz} \tag{14}$$

Eqn. (14) provides the coefficients $a_y$, $a_{xy}$, $a_z$ and $a_{xz}$ that are needed in eqn. (7). Analogous terms in eqns. (8) and (9) can now be made via a cyclic rotation of variables. Thus one cyclic rotation of variables applied to eqn. (14) provides us $b_z$, $b_{yz}$, $b_x$ and $b_{xy}$. Another such rotation of variables yields $c_x$, $c_{xz}$, $c_y$ and $c_{yz}$. The constraint applied to the quadratic terms in eqns. (7) to (9) gives:

$$a_{xx} = -\frac{1}{2}(b_{xy} + c_{xz}) - \frac{3}{35} a_{xxxx} - \frac{1}{20}(b_{xyyy} + c_{xzzz}) \quad (15)$$

Analogous terms in eqns. (8) and (9) can now be made by applying cyclic rotations to variables in eqn. (15) and those rotations yield $b_{yy}$ and $c_{zz}$.

Matching the constant terms at the $x = \pm 1/2$ boundaries gives:

$$a_0 = \frac{1}{2}\left(B_0^{x+} + B_0^{x-}\right) - \frac{1}{6} a_{xx} - \frac{1}{70} a_{xxxx} \quad ; a_x = \left(B_0^{x+} - B_0^{x-}\right) - \frac{1}{10} a_{xxx} \quad (16)$$

Eqn. (1) provides the coefficients $a_0$ and $a_x$ that are needed in eqn. (7). Analogous terms in eqns. (8) and (9) can now be made to get $b_0$, $b_y$, $c_0$ and $c_z$. The constraint applied to the linear terms in eqns. (7) to (9) gives:

$$\left(a_x + b_y + c_z\right) + \frac{1}{10}\left(a_{xxx} + b_{yyy} + c_{zzz}\right) = 0 \quad (17)$$

The coefficients in eqn. (16) are so constructed that, along with eqn. (17), they ensure (and are equivalent to) the integral form of the divergence-free constraint:

$$\left(B_0^{x+} - B_0^{x-}\right) + \left(B_0^{y+} - B_0^{y-}\right) + \left(B_0^{z+} - B_0^{z-}\right) = 0 \quad (18)$$

This completes our description of the divergence-free reconstruction on the unit cube.

In practical situations, one might want to carry out the same procedure on a zone of size $\Delta x$, $\Delta y$ and $\Delta z$ in the x, y and z-directions respectively. Notice that eqn. (18) then becomes:

$$\frac{1}{\Delta x}\left(B_0^{x+} - B_0^{x-}\right) + \frac{1}{\Delta y}\left(B_0^{y+} - B_0^{y-}\right) + \frac{1}{\Delta z}\left(B_0^{z+} - B_0^{z-}\right) = 0 \qquad (19)$$

The problem can be mapped to a unit cube by dividing all the coefficients in eqns. (4), (5) and (6) by $\Delta x$, $\Delta y$ and $\Delta z$ respectively. The method described in this Section can now be applied to get the coefficients in eqns. (7), (8) and (9) and all the coefficients in those equations can subsequently be multiplied by $\Delta x$, $\Delta y$ and $\Delta z$ respectively. This completes our description of the divergence-free reconstruction on any rectilinear mesh.

We make a few observations below:

1) We observe that the normal components of the magnetic field in eqns. (4) to (6) are indeed fourth order accurate in the faces. Furthermore, specifying all the moments in eqns. (4) to (6) at the zone faces uniquely specifies all the coefficients in eqns. (7) to (9) for the interior of that zone. Eqns. (7) to (9) contain all the terms that one would need in a fourth order accurate polynomial expansion. Thus all the fourth order accurate terms that are needed for reconstructing a divergence-free vector field in the interior of a zone are already provided by their fourth order accurate specification at the boundaries. The few remaining terms in eqns. (7) to (9) only help in matching the magnetic fields exactly to the components at the boundaries. By dropping suitable terms in eqns. (4) to (9) we can also see that all the third order accurate terms that are needed for reconstructing a divergence-free vector field in the interior of a zone are already provided by their third order accurate specification at the boundaries. A similar statement applies to the second order accurate reconstruction.

2) Notice too that when carrying out adaptive mesh refinement of a divergence-free vector field by a refinement ratio of three, we need to specify nine degrees of freedom at each boundary. The fourth order reconstruction presented here has ten degrees of freedom at each boundary, see eqns. (4) to (6). One degree of freedom can be relinquished either by setting $B_{yyy}^{x\pm} = B_{zzz}^{x\pm}$ or by setting $B_{yyz}^{x\pm} = B_{yzz}^{x\pm}$. Thus the reconstruction has sufficient amount of freedom to make it useful for carrying out adaptive mesh refinement with refinement ratios of three.

3) Balsara [3] provided formulae for carrying out adaptive mesh refinement of a divergence-free vector field by a refinement ratio of two. The above point shows that a refinement ratio of three is also easy to achieve. Recursive application of the algorithms makes it possible to achieve refinement ratios that are any multiples or two and three. The algorithm presented here is dimensionally unsplit and offers analytic, closed form expressions for the reconstruction. Our formulation also minimizes the energy of the magnetic field and we will later show in Section 5 that it helps keep the pressure positive when simulating stringent test problems.

4) The same transformations that were catalogued in Balsara [4] for treating logically rectangular meshes with diagonal metrics go over transparently for the reconstruction given here. As a result, there are no obstacles to using the present formulation for designing MHD algorithms in cylindrical and spherical meshes. Similarly, one can use the present formulation for carrying out adaptive mesh refinement on such curvilinear meshes.

5) The present formulation should also help in making divergence-free prolongation which is very useful in the construction of divergence-free multigrid schemes for resistive or Hall MHD.

## 3) Step-by-Step Description of the RK-WENO Schemes for Divergence-free MHD

The equations of ideal MHD can be cast in a conservative form that is suited for the design of higher order Godunov schemes. In that form they become:

$$\frac{\partial \mathbf{U}}{\partial t} + \frac{\partial \mathbf{F}}{\partial x} + \frac{\partial \mathbf{G}}{\partial y} + \frac{\partial \mathbf{H}}{\partial z} = 0 \qquad (20)$$

where $\mathbf{F}$, $\mathbf{G}$ and $\mathbf{H}$ are the ideal fluxes. Written out explicitly, eqn. (20) becomes:

$$\frac{\partial}{\partial t}\begin{pmatrix} \rho \\ \rho v_x \\ \rho v_y \\ \rho v_z \\ \varepsilon \\ B_x \\ B_y \\ B_z \end{pmatrix} + \frac{\partial}{\partial x}\begin{pmatrix} \rho v_x \\ \rho v_x^2 + P + \mathbf{B}^2/8\pi - B_x^2/4\pi \\ \rho v_x v_y - B_x B_y/4\pi \\ \rho v_x v_z - B_x B_z/4\pi \\ (\varepsilon+P+\mathbf{B}^2/8\pi)v_x - B_x(\mathbf{v}\cdot\mathbf{B})/4\pi \\ 0 \\ (v_x B_y - v_y B_x) \\ -(v_z B_x - v_x B_z) \end{pmatrix}$$

$$+ \frac{\partial}{\partial y}\begin{pmatrix} \rho v_y \\ \rho v_x v_y - B_x B_y/4\pi \\ \rho v_y^2 + P + \mathbf{B}^2/8\pi - B_y^2/4\pi \\ \rho v_y v_z - B_y B_z/4\pi \\ (\varepsilon+P+\mathbf{B}^2/8\pi)v_y - B_y(\mathbf{v}\cdot\mathbf{B})/4\pi \\ -(v_x B_y - v_y B_x) \\ 0 \\ (v_y B_z - v_z B_y) \end{pmatrix} + \frac{\partial}{\partial z}\begin{pmatrix} \rho v_z \\ \rho v_x v_z - B_x B_z/4\pi \\ \rho v_y v_z - B_y B_z/4\pi \\ \rho v_z^2 + P + \mathbf{B}^2/8\pi - B_z^2/4\pi \\ (\varepsilon+P+\mathbf{B}^2/8\pi)v_z - B_z(\mathbf{v}\cdot\mathbf{B})/4\pi \\ (v_z B_x - v_x B_z) \\ -(v_y B_z - v_z B_y) \\ 0 \end{pmatrix} = 0$$

$$(21)$$

where $\rho$ is the density, $v_x$, $v_y$ and $v_z$ are the velocity components, $B_x$, $B_y$ and $B_z$ are the magnetic field components, $\gamma$ is the adiabatic index and $\varepsilon = \rho v^2/2 + P/(\gamma - 1) + \mathbf{B}^2/8\pi$ is the total energy. The equations for the density, momentum density and energy density parallel those in the Euler equations and can be discretized using standard RKDG formulations. While the magnetic fields seem to have a conservation law structure, an examination of the flux vectors show that the equations of MHD obey the following symmetries:

$$F_7 = -G_6, \quad F_8 = -H_6, \quad G_8 = -H_7 \tag{22}$$

These symmetries are also obeyed when any manner of non-ideal terms are introduced and are a fundamental consequence of the induction equation, see eqn. (1). Balsara & Spicer [10] realized how to use this dualism between the flux components and the electric fields to build electric fields at zone edges using the properly upwinded Godunov fluxes. Balsara [4] introduced a better way of obtaining the electric fields at zone edges that avoids spatial averaging. The Balsara & Spicer [10] scheme is inherently second order accurate because of the spatial averaging. By overcoming this limitation, the Balsara [4] scheme is easily extended to all orders. Once the electric fields are obtained at requisite collocation points on the zone edges a discrete version of eqn. (1) can be built, as shown in Balsara [4]. Balsara [4] also showed that Runge-Kutta time-discretizations could be used for MHD. We therefore describe the steps in the implementation of a Runge-Kutta time-discretiztion for MHD. The spatial representation is provided by an efficient implementation of a WENO scheme for structured meshes. A step-by-step description of the WENO scheme with Runge-Kutta time-stepping is provided below.

**3.1) Divergence-Free WENO Reconstruction Step**

The first step in any finite volume scheme consists of obtaining a reconstruction of the field variables within a zone. Inclusion of the appropriate moments of the flow yields a correspondingly high accuracy. Thus at any stage in a multi-stage RK time-stepping scheme our first task is to obtain a representation of the flow in the following basis space:

$$\begin{aligned}
\mathbf{U}(x,y,z) = {} & \bar{\mathbf{U}}_1 P_0(x) P_0(y) P_0(z) \\
& + \hat{\mathbf{U}}_2 P_1(x) P_0(y) P_0(z) + \hat{\mathbf{U}}_3 P_0(x) P_1(y) P_0(z) + \hat{\mathbf{U}}_4 P_0(x) P_0(y) P_1(z) \quad \leftarrow \text{ second order} \\
& + \hat{\mathbf{U}}_5 P_2(x) P_0(y) P_0(z) + \hat{\mathbf{U}}_6 P_0(x) P_2(y) P_0(z) + \hat{\mathbf{U}}_7 P_0(x) P_0(y) P_2(z) \\
& + \hat{\mathbf{U}}_8 P_1(x) P_1(y) P_0(z) + \hat{\mathbf{U}}_9 P_0(x) P_1(y) P_1(z) + \hat{\mathbf{U}}_{10} P_1(x) P_0(y) P_1(z) \quad \leftarrow \text{ third order} \\
& + \hat{\mathbf{U}}_{11} P_3(x) P_0(y) P_0(z) + \hat{\mathbf{U}}_{12} P_0(x) P_3(y) P_0(z) + \hat{\mathbf{U}}_{13} P_0(x) P_0(y) P_3(z) \\
& + \hat{\mathbf{U}}_{14} P_2(x) P_1(y) P_0(z) + \hat{\mathbf{U}}_{15} P_2(x) P_0(y) P_1(z) + \hat{\mathbf{U}}_{16} P_1(x) P_2(y) P_0(z) + \hat{\mathbf{U}}_{17} P_0(x) P_2(y) P_1(z) \\
& + \hat{\mathbf{U}}_{18} P_1(x) P_0(y) P_2(z) + \hat{\mathbf{U}}_{19} P_0(x) P_1(y) P_2(z) + \hat{\mathbf{U}}_{20} P_1(x) P_1(y) P_1(z) \quad \leftarrow \text{ fourth order}
\end{aligned}$$

(23)

Here (x,y,z) denotes the local coordinates in the unit cube $[-1/2,1/2]\times[-1/2,1/2]\times[-1/2,1/2]$ to which the zone of interest is mapped and $\mathbf{U}(x,y,z)$ is the vector of conserved variables from eqn. (20). $\hat{\mathbf{U}}_{2,...,20}$ are the modes that are reconstructed at each time level for a fourth order scheme, with fewer modes needed for lower order schemes. The first five components of $\bar{\mathbf{U}}_1$ are just the zone-averaged mass, momentum and total energy densities that are available in each zone. The last three components of $\bar{\mathbf{U}}_1$ and $\hat{\mathbf{U}}_{2,...,20}$ have to be obtained from the divergence-free reconstruction of the magnetic field, whose facially-averaged components are available at the appropriate faces. Using WENO reconstruction in each of the faces we obtain all the moments of eqns. (4) to (6). The results of Section 2 then gives us all the moments of eqns. (7) to (9) which also gives us the last three components of $\bar{\mathbf{U}}_1$ and $\hat{\mathbf{U}}_{2,...,20}$. WENO reconstruction can now be applied to obtain all the remaining components of $\hat{\mathbf{U}}_{2,...,20}$. Several good choices are available for WENO interpolation these days including the works of Jiang & Shu [25], Balsara & Shu[8], Dumbser & Käser [21], Balsara et al [5] and [7]. In Balsara et al [7] we presented a WENO reconstruction strategy that is very well-suited for structured meshes and we used that strategy here.

**3.2) Flux and Electric Field Evaluation Step**

A higher order scheme should also evaluate fluxes and electric fields with suitably high accuracy. Traditionally this has been done by solving a large number of Riemann problems at a large number of quadrature points as was done in Cockburn & Shu [14]. A substantially simpler strategy was presented in Dumbser, Enaux & Toro [20] where the flux is viewed as a linear combination of four vectors. The four vectors are: a) the conserved variables to the left of the zone boundary given by $\mathbf{U}_{L;\,i+1/2,j,k}(y,z)$, b) the conserved variables to the right of the zone boundary given by $\mathbf{U}_{R;\,i+1/2,j,k}(y,z)$, c) the flux to the left of the zone boundary given by $\mathbf{F}_{L;\,i+1/2,j,k}(y,z)$ and d) the flux to the right of the zone boundary given by $\mathbf{F}_{R;\,i+1/2,j,k}(y,z)$. The strategy proposed by Dumbser, Enaux & Toro [20] applies to the space-time domain. We specialize it for the case where the time-averaging is not needed. Below it is instantiated for the linearized Riemann solver at any general point (y,z) on the x-boundary "*i+1/2,j,k*". Such a flux is described by:

$$\mathbf{F}_{i+1/2,j,k}(y,z) = \frac{1}{2}\left(\mathbf{F}_{L;\,i+1/2,j,k}(y,z) + \mathbf{F}_{R;\,i+1/2,j,k}(y,z)\right) - \left|A(y,z)\right|\left(\mathbf{U}_{R;\,i+1/2,j,k}(y,z) - \mathbf{U}_{L;\,i+1/2,j,k}(y,z)\right)$$

(24)

As written, the matrix $\left|A(y,z)\right|$ would have to be evaluated anew at each point (x,y) on the zone boundary. The essential insight from Dumbser, Enaux & Toro [20] consists of realizing that $\left|A(y,z)\right|$ can be evaluated once at the barycenter of the zone boundary. This is equivalent to freezing the dissipation model all over the zone boundary and it also makes the flux a linear function of the four vectors catalogued above. $\mathbf{U}_{L;\,i+1/2,j,k}(y,z)$ and $\mathbf{U}_{R;\,i+1/2,j,k}(y,z)$ are easily obtained once the reconstruction from eqn. (23) is available in the two zones that abut a zone face. Balsara et al [7] present a very efficient strategy for obtaining $\mathbf{F}(x,y,z)$ within a zone when eqn. (23) is available in the zone. As a result, $\mathbf{F}_{L;\,i+1/2,j,k}(y,z)$ and $\mathbf{F}_{R;\,i+1/2,j,k}(y,z)$ are also easily obtained. Averaging eqn. (24) over the (y,z) coordinates of an x-face of the reference element only entails evaluating the integral analytically once and is easily done by using a symbolic manipulation package. A similar strategy can be applied at the y and z-faces. The electric fields are also easily obtained by averaging eqn. (24) suitably over the edges of the reference element and picking out the appropriate components of the fluxes. Four electric field contributions are available at each edge, one from each of the four faces that come together at that edge. These electric fields are averaged, as in Balsara [4] to obtain the final electric field at the zone of interest. This completes our description of the fluid flux and the electric field evaluation for any stage in our multi-stage RK time-update.

**3.3) Multi-Stage Runge Kutta Time Update Step**

The strong stability preserving Runge Kutta schemes from Shu & Osher [31] and [32] are used for carrying out a time update. At each stage of the multi-stage RK update, we apply the steps from Sub-Sections 3.1 and 3.2 to obtain the fluxes at each face and the electric field components at each edge. The Runge-Kutta time-stepping schemes consist of writing eqn. (1) for the magnetic field evolution and eqn. (20) for the evolution of the mass, momentum and energy densities in the form

$$\frac{d\mathbf{U}}{dt} = L(\mathbf{U}) \tag{25}$$

Where $L(\mathbf{U})$ is a discretization of the spatial operator. The second order TVD Runge-Kutta scheme is simply the Heun scheme:

$$\begin{aligned} \mathbf{U}^{(1)} &= \mathbf{U}^n + \frac{1}{2} \Delta t\, L(\mathbf{U}^n) \\ \mathbf{U}^{n+1} &= \mathbf{U}^n + \Delta t\, L(\mathbf{U}^{(1)}) \end{aligned} \tag{26}$$

The third order TVD Runge-Kutta scheme is given by:

$$\mathbf{U}^{(1)} = \mathbf{U}^n + \Delta t\, L(\mathbf{U}^n)$$

$$\mathbf{U}^{(2)} = \frac{3}{4}\mathbf{U}^n + \frac{1}{4}\mathbf{U}^{(1)} + \frac{1}{4}\Delta t\, L(\mathbf{U}^{(1)}) \qquad (27)$$

$$\mathbf{U}^{n+1} = \frac{1}{3}\mathbf{U}^n + \frac{2}{3}\mathbf{U}^{(2)} + \frac{2}{3}\Delta t\, L(\mathbf{U}^{(2)})$$

The fourth order RK scheme from Shu & Osher [31] is rather complicated to implement and was not implemented here. As a result, the temporal update of the spatially fourth order scheme was always done with eqn. (27). For most applications this yields a serviceable scheme that functions at a robust Courant number. However when demonstrating the order of accuracy at fourth order in Section 4 we had to reduce the Courant number by a factor of ~ 0.396 for every doubling of the number of zones. This had to be done so that the third order temporal accuracy from eqn. (27) keeps step with the fourth order spatial accuracy. This deficiency is ameliorated by the ADER (for Arbitrary Derivative Riemann Problem) schemes presented in Balsara et al [7].

## 4) Accuracy Analysis

The schemes presented here handily meet their design accuracies in one dimension. It is therefore interesting to present multi-dimensional tests showing high order of accuracy. Here we present a couple of demonstrations of high accuracy in two and three dimensions. A more extensive accuracy analysis for hydrodynamic and MHD problems has been catalogued in Balsara et al [7] for a new class of ADER-WENO schemes.

A couple of points need to be made about the simulations presented here. First, following Balsara [4] we used the slopes from the r=3 WENO reconstruction of Jiang & Shu [25] for our second order scheme. As a result, the slopes have one more order of accuracy than the accuracy that would be furnished by a TVD-preserving limiter. This yields a very superior second order scheme. Second, for all the accuracy analyses presented in this section involving the spatially fourth order scheme, the Courant number was always decreased by a factor of 0.396 for every doubling of the number of zones.

**4.1) Magnetized Isodensity Vortex in Two Dimensions**

This test problem as described in Balsara [4] consists of a magnetized vortex moving across a domain given by [-5, 5] x [-5, 5] at an angle of 45° for a time of 10 units. For the fourth order scheme the domain is increased to [-10, 10] x [-10, 10] and the simulation time is increased to 20 units. This is done because the magnetic field has a Gaussian decay with radius and the smaller domain retains a small but significant amount of magnetic field at the boundary. Had we used the smaller domain for the fourth order scheme, this small but spurious magnetic field would actually have been picked up by the scheme and its order property would have been damaged. The problem is initialized with an unperturbed flow of $(\rho, P, v_x, v_y, B_x, B_y) = (1, 1, 1, 1, 0, 0)$. All boundaries are periodic. The ratio of the specific heat is set to $\gamma = 5/3$. The vortex is set up as a fluctuation of the unperturbed flow in the velocities and the magnetic field given by:

$$(\delta v_x, \delta v_y) = \frac{\kappa}{2\pi} e^{0.5(1-r^2)}(-y, x)$$

$$(\delta B_x, \delta B_y) = \frac{\mu}{2\pi} e^{0.5(1-r^2)}(-y, x)$$

The pressure fluctuation can be written as

$$\delta P = \frac{1}{8\pi}(\frac{\mu}{2\pi})^2 (1-r^2) e^{(1-r^2)} - \frac{1}{2}(\frac{\kappa}{2\pi})^2 e^{(1-r^2)}$$

The density is set to unity. A Courant number of 0.4 was used for all the second and third order test problems and also for the coarsest mesh in the fourth order test problem. A linearized Riemann solver was used.

TABLE I

| Method | Number of zones | $L_1$ error | $L_1$ order | $L_\infty$ error | $L_\infty$ order |
|---|---|---|---|---|---|
| 2nd order ADER | 32×32 | $1.15689 \times 10^{-2}$ | | 0.189318 | |
| | 64×64 | $3.74953 \times 10^{-3}$ | 1.62 | $6.00319 \times 10^{-2}$ | 1.66 |
| | 128×128 | $9.57467 \times 10^{-4}$ | 1.97 | $1.53503 \times 10^{-2}$ | 1.97 |
| | 256×256 | $2.39584 \times 10^{-4}$ | 2.00 | $3.83531 \times 10^{-3}$ | 2.00 |
| 3rd order ADER | 32×32 | $5.53837 \times 10^{-3}$ | | $9.79331 \times 10^{-2}$ | |
| | 64×64 | $9.77841 \times 10^{-4}$ | 2.50 | $1.75191 \times 10^{-2}$ | 2.48 |
| | 128×128 | $1.27506 \times 10^{-4}$ | 2.94 | $2.36221 \times 10^{-3}$ | 2.89 |
| | 256×256 | $1.60549 \times 10^{-5}$ | 2.99 | $2.99136 \times 10^{-4}$ | 2.98 |
| 4th order ADER | 32×32 | $2.96778 \times 10^{-3}$ | | 0.103623 | |
| | 64×64 | $1.56211 \times 10^{-4}$ | 4.25 | $5.21875 \times 10^{-3}$ | 4.31 |
| | 128×128 | $7.33125 \times 10^{-6}$ | 4.41 | $2.45447 \times 10^{-4}$ | 4.41 |

Table I shows the results of the accuracy analysis. The error is measured in the x-component of the magnetic field. All the schemes meet their design accuracies. Notice that the third order scheme at 128x128 zone resolution shows the same $L_1$ error as the second order scheme at 256x256 zone resolution. We see therefore that higher order schemes deliver a much improved solution quality compared to lower order schemes on meshes of the same resolution. Furthermore the higher order schemes need far fewer zones to achieve the same accuracy as a lower order scheme. Table I therefore illustrates the utility of higher order schemes very nicely.

**4.2) Torsional Alfven Wave Propagation in Three Dimensions**

The previous test problem used a flow that was an exact, equilibrium structure of the governing equations. Although torsional Alfven waves also satisfy the governing equations, they are susceptible to parametric instabilities. These instabilities exist at low values of plasma-β , see Goldstein [23] and Del Zanna et al [18], and also at high values of plasma-β , see Jayanti & Hollweg [24]. The present test problem is designed to ameliorate such instabilities as far as possible.

In this problem we initialize a torsional Alfven wave along the $x'$ axis of an ($x'$, $y'$, $z'$) coordinate system with the following parameters

$$\rho = 1 \; , P = 1000 \; , \Phi = \frac{2\pi}{\lambda}\left(x' - 2\,t\right)$$

$$v_{x'} = 1 \; , v_{y'} = \varepsilon \cos \Phi \; , v_{z'} = \varepsilon \sin \Phi$$

$$B_{x'} = \sqrt{4\pi\rho} \; , B_{y'} = -\varepsilon\sqrt{4\pi\rho} \cos \Phi \; , B_{z'} = -\varepsilon\sqrt{4\pi\rho} \sin \Phi$$

Here we take $\varepsilon = 0.02$ and $\lambda = \sqrt{3}$. The magnetic vector potential is also useful when initializing a divergence-free magnetic field on a mesh and is given by

$$A_{x'} = 0 \; , A_{y'} = \varepsilon\lambda\sqrt{\rho/\pi} \cos \Phi \; , A_{z'} = \sqrt{4\pi\rho} \; y' + \varepsilon\lambda\sqrt{\rho/\pi} \sin \Phi$$

The actual problem is solved on a unit cube in the (x,y,z) coordinate frame which is rotated relative to the (x', y', z') coordinate system. The rotation matrix is called **A** and is given by

$$\mathbf{A} = \begin{bmatrix} \cos\psi \cos\phi - \cos\theta \sin\phi \sin\psi & \cos\psi \sin\phi + \cos\theta \cos\phi \sin\psi & \sin\psi \sin\theta \\ -\sin\psi \cos\phi - \cos\theta \sin\phi \cos\psi & -\sin\psi \sin\phi + \cos\theta \cos\phi \cos\psi & \cos\psi \sin\theta \\ \sin\theta \sin\phi & -\sin\theta \cos\phi & \cos\theta \end{bmatrix}$$

where $\phi = -\pi/4$, $\theta = \sin^{-1}\left(-\sqrt{2/3}\right)$ and $\psi = \sin^{-1}\left(\left(\sqrt{2}-\sqrt{6}\right)/4\right)$. As a result, the position vector **r'** in the primed frame transforms to the position vector **r** in the unprimed frame as $\mathbf{r} = \mathbf{A}\,\mathbf{r'}$. Other vectors transform similarly. Application of the rotation matrix makes the wave propagate along the diagonal of the unit cube. The wave propagates at a speed of 2 units. The problem is stopped at a time of $\sqrt{3}/2$ by which time the wave has propagated once around the unit cube. A Courant number of 0.3 was used for all the second and third order test problems and also for the coarsest mesh in the fourth order test problem. A linearized Riemann solver was used.

Table II

| Method | Number of zones | $L_1$ error | $L_1$ order | $L_\infty$ error | $L_\infty$ order |
|---|---|---|---|---|---|
| 2nd order ADER CG | 8×8×8 | 3.46827 × 10$^{-2}$ | | 5.17569 × 10$^{-2}$ | |
| | 16×16×16 | $2.25885 \times 10^{-2}$ | 0.62 | $3.57951 \times 10^{-2}$ | 0.53 |
| | 32×32×32 | $4.87419 \times 10^{-3}$ | 2.21 | $7.68322 \times 10^{-3}$ | 2.22 |
| | 48×48×48 | $1.77966 \times 10^{-3}$ | 2.48 | $2.79747 \times 10^{-3}$ | 2.49 |
| 3$^{rd}$ order ADER CG | 8×8×8 | $3.56043 \times 10^{-2}$ | | $5.32694 \times 10^{-2}$ | |
| | 16×16×16 | $1.65967 \times 10^{-2}$ | 1.10 | $2.56119 \times 10^{-2}$ | 1.06 |
| | 32×32×32 | $2.65506 \times 10^{-3}$ | 2.64 | $4.17435 \times 10^{-3}$ | 2.62 |
| | 48×48×48 | $8.05482 \times 10^{-4}$ | 2.94 | $1.27225 \times 10^{-3}$ | 2.93 |
| 4$^{th}$ order ADER CG | 8×8×8 | $2.52284 \times 10^{-2}$ | | $3.82295 \times 10^{-2}$ | |
| | 16×16×16 | $1.17975 \times 10^{-3}$ | 4.42 | $1.85115 \times 10^{-3}$ | 4.37 |
| | 32×32×32 | $5.29206 \times 10^{-5}$ | 4.48 | $8.38025 \times 10^{-5}$ | 4.47 |

Table II presents the accuracy analysis for schemes up to fourth order. Please recall that the combination of a spatially fourth order scheme with a temporally third order RK scheme required us to use a diminishing Courant number with increasing resolution at fourth order and only at fourth order. As a result, the accuracy analysis of the fourth order scheme had to be restricted to smaller meshes. In Balsara et al [7] we present schemes that overcome this limitation. Table II is nevertheless very illustrative. It shows that all the schemes presented here meet their design accuracies. We see that even on very small resolution starved meshes, such as the 16x16x16 mesh in Table I, the fourth order scheme offers more than one order of magnitude improvement over the second order scheme. Table II therefore provides a further illustration of the utility of higher order schemes.

## 5) Test Problems

In this section we present several tests for the schemes that have been designed here. Because the divergence-free reconstruction of the magnetic field only comes to the fore in multiple dimensions, all of the tests presented here are inherently two-dimensional and were run with a Courant number of 0.4.

### 5.1) Numerical Dissipation and Long-Term Decay of Alfven Waves in Two Dimensions

This test problem was first presented in Balsara [4] and examines the dissipation of torsional Alfven waves in two dimensions. Here the torsional Alfven waves propagate at an angle of $\tan^{-1}(1/r) = \tan^{-1}(1/6) = 9.462°$ to the y-axis through a domain given by [-r/2, r/2] x [-r/2, r/2] with r = 6. The problem was initialized on a computaitonal domain with 120 x 120 zones. Periodic boundary conditions were enforced. The pressure and density are uniformly initialized as $P_0 = 1$ and $\rho_0 = 1$. The unperturbed velocity and unperturbed magnetic field are given by $v_0 = 0$ and $B_0 = 1$. The amplitude of the Alfven waves is parametrized by $\varepsilon$, which is set to 0.2. The simulation was stopped at 129 time units by which time the waves had crossed the domain several times. The CFL number was set to 0.4 for all the schemes presented here. The direction of the wave propagation along the unit vector can be written as

$$\hat{n} = n_x \hat{i} + n_y \hat{j} = \frac{1}{\sqrt{r^2+1}} \hat{i} + \frac{r}{\sqrt{r^2+1}} \hat{j}.$$

The phase of the waves is given by

$$\phi = \frac{2\pi}{n_y}(n_x x + n_y y - V_A t), \text{ where } V_A = \frac{B_0}{\sqrt{4\pi\rho_0}}.$$

The velocity is given by

$$\mathbf{v} = (v_0 n_x - \varepsilon n_y \cos\phi)\hat{i} + (v_0 n_y - \varepsilon n_x \cos\phi)\hat{j} + \varepsilon \sin\phi \hat{k}.$$

The magnetic field is given by

$$\mathbf{B} = (B_0 n_x + \varepsilon n_y \sqrt{4\pi\rho_0} \cos\phi)\hat{i} + (B_0 n_y - \varepsilon n_x \sqrt{4\pi\rho_0} \cos\phi)\hat{j} - \varepsilon\sqrt{4\pi\rho_0} \sin\phi \hat{k}.$$

The corresponding vector potential is given by

$$\mathbf{A} = -\frac{\varepsilon\sqrt{4\pi\rho_0}}{2\pi}\cos\phi\hat{i} + (-B_0 n_y x + B_0 n_x y + \frac{\varepsilon n_y \sqrt{4\pi\rho_0}}{2\pi}\sin\phi)\hat{k}$$

and is used to initialize the magnetic field.

Fig. 1 shows the time-evolution of the maximum of the z-velocity and the maximum of the z-component of magnetic field. All the panels in Fig. 1 use log-linear scaling. To explore the effect of Riemann solvers on this problem, the HLL and linearized Riemann solvers were used with the second, third and fourth order schemes. For comparison purposes, we also present results from a second order TVD scheme using vanLeer's MC limiter. We see that regardless of the Riemann solver used, increasing the order of accuracy provides a substantial reduction in the numerical dissipation. Thus higher order schemes are favored for the simulation of complex phenomena involving wave propagation. For the lower order schemes the linearized Riemann solver offers a significant improvement over the HLL Riemann solver. However, this advantage is diminished with increasing order. We therefore see that higher order schemes allow us to get by with less expensive Riemann solvers.

**5.2) The Rotor Problem in Two Dimensions**

The two dimensional rotor problem was presented in Balsara & Spicer [10] and in Balsara [4]. The description in Balsara [4] is quite thorough. As a result the problem description is not repeated here. As in Balsara [4] the problem was set up on a 200x200 zone mesh and was run with a Courant number of 0.4 to a completion time of 0.29 units. The spatially fourth order WENO scheme with a third order RK time-stepping strategy and a linearized Riemann solver were used. Fig. 2 shows the density, pressure, Mach number and the magnitude of the magnetic field. The results are very consistent with those from Balsara & Spicer [10] showing that the divergence-free reconstruction presented here performs well on multi-dimensional MHD problems.

**5.3) The Blast Problem in Two Dimensions**

This two-dimensional problem was first presented by Balsara & Spicer [10]. It has also been catalogued in detail in Balsara [4] and we do not repeat the same description here. The fourth order WENO scheme with a third order RK time-stepping strategy and an HLL Riemann solver was applied to a mesh having 200 × 200 zones. The problem was run with a Courant number of 0.4 and was stopped at a time of 0.01 units. The problem results in an extremely strong, almost circular fast magnetosonic shock propagating at all possible angles to the magnetic field in the low-β ambient plasma. The plasma-β is 0.000251 making this a challenging test problem. Fig. 3 shows the logarithm (base 10) of the density, the logarithm of the pressure, the magnitude of the velocity and the magnitude of the magnetic field. We see that all structures are captured crisply. The positivity of the pressure is maintained even in regions where the strong shock propagates obliquely to the mesh. This shows that the divergence-free reconstruction strategies and the resultant high order schemes presented here perform well on stringent multi-dimensional MHD problems involving low-β plasmas.

## 6) Conclusions

The work presented here enables us to come to the following conclusions:

1) Following a line of development begun in Balsara [3], we show that the problem of reconstructing divergence-free vector fields can be carried out to higher orders.

2) Following a line of development begun in Balsara [4], we show that the above development yields divergence-free WENO schemes with order of accuracy that is better than second. In particular, we explore the third and fourth order accurate schemes here.

3) When applied to smooth test problems, the schemes have been shown to meet their design accuracies.

4) Using a stringent set of test problems we show that the schemes presented here effectively combine the dual, and often-conflicting demands of capturing very strong shocks and retaining low dissipation in contact discontinuities and Alfven waves. This shows the effectiveness of our schemes for numerical MHD.

## Acknowledgements

DSB acknowledges support via NSF grant AST-0607731. DSB also acknowledges NASA grants HST-AR-10934.01-A, NASA-NNX07AG93G and NASA-NNX08AG69G. The majority of simulations were performed on PC clusters at UND but a few initial simulations were also performed at NASA-NCCS.

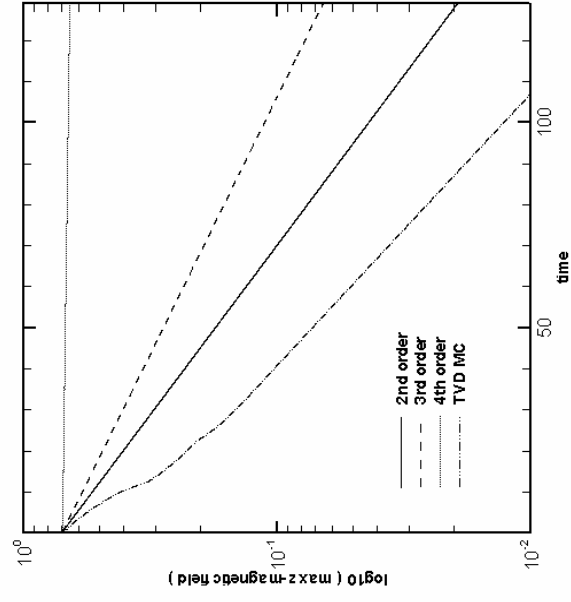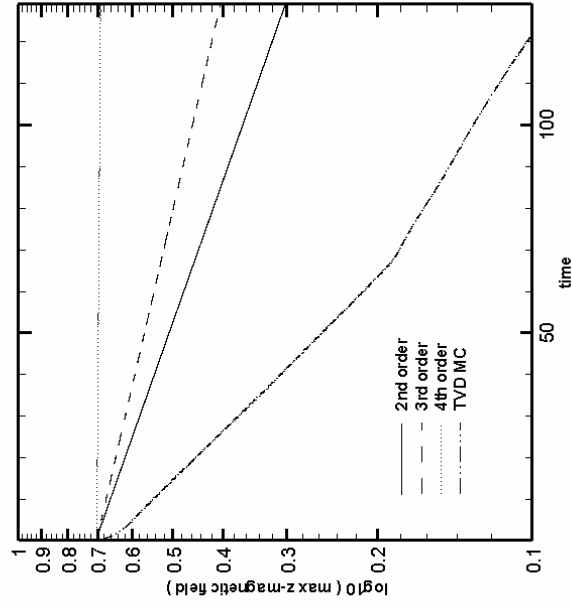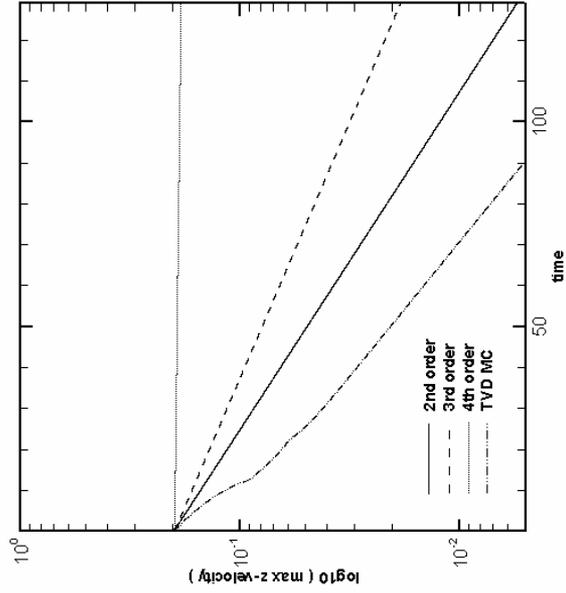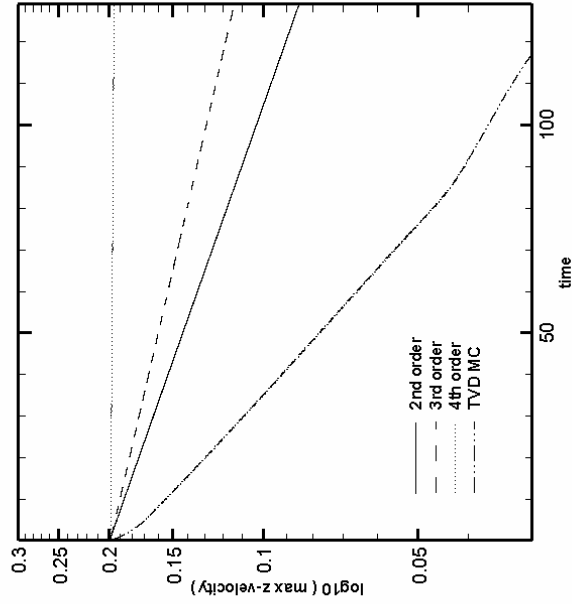

Fig 1) The log-linear plots show the decay of torsional Alfven waves that propagate obliquely on a two dimensional square. The top two panels show the decay of the maximum z-velocity and the maximum z-component of the magnetic field when second, third and fourth order schemes are used with an HLLE Riemann solver. The bottom two panels show the same information when a linearized Riemann solver is used. For comparison purposes, the results from a TVD scheme with MC limiter are also shown. Temporally third order RK updates were used for the spatially fourth order scheme. Observe that the decay is substantially reduced with increasing spatial order. Also observe that the linearized Riemann solver provides a substantial improvement to the solution, especially at lower orders.

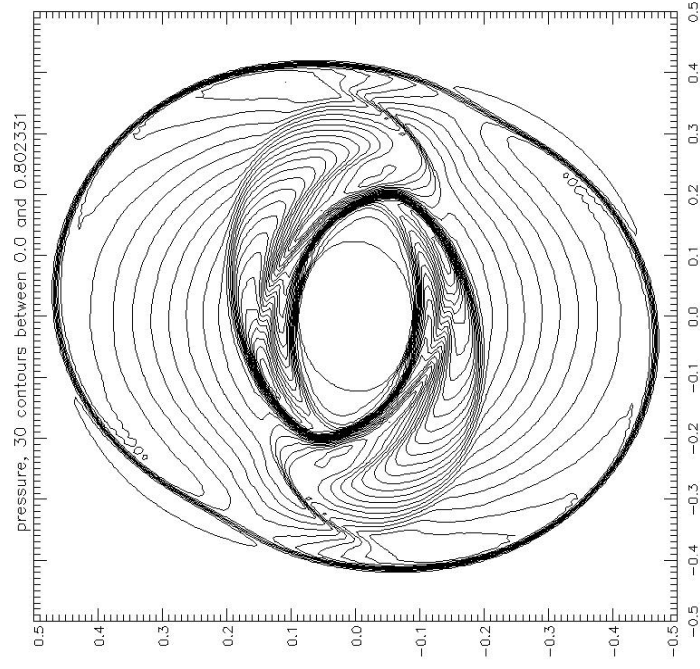
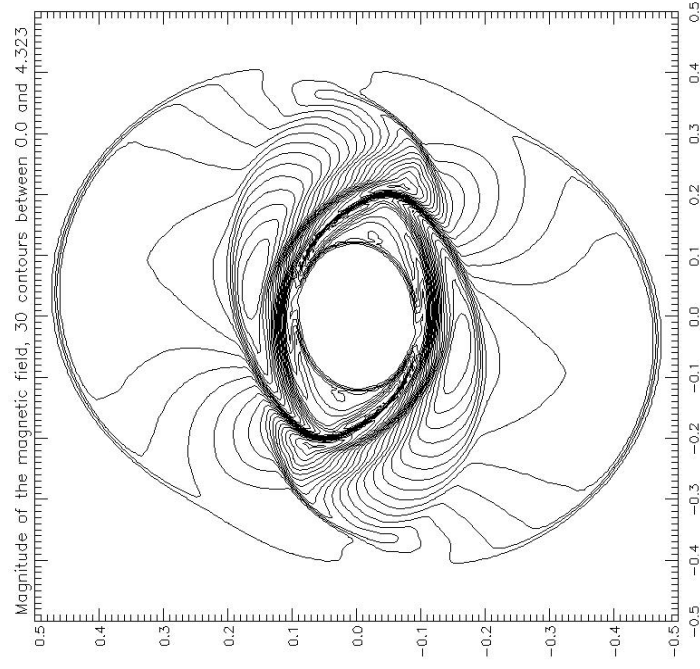
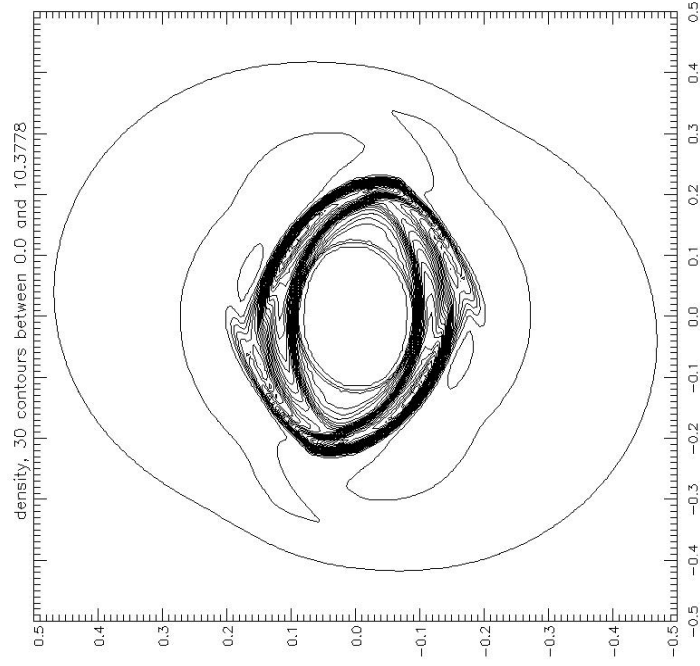
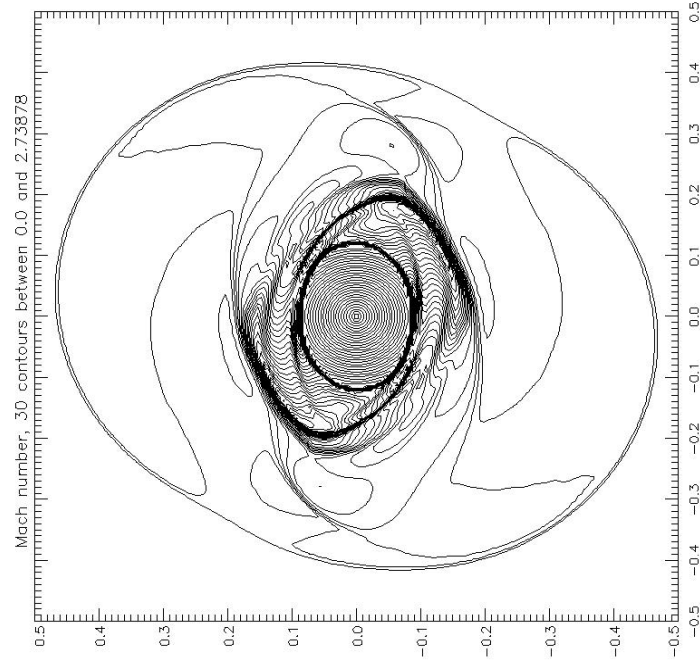

Fig. 2) Rotor problem where the top two panels show the density and pressure and the bottom two panels show the Mach number and the magnitude of the magnetic field. The fourth order WENO scheme with third order RK timestepping and a linearized Riemann solver was used. 30 contours are shown for each figure with the min and max values catalogued above the panels.

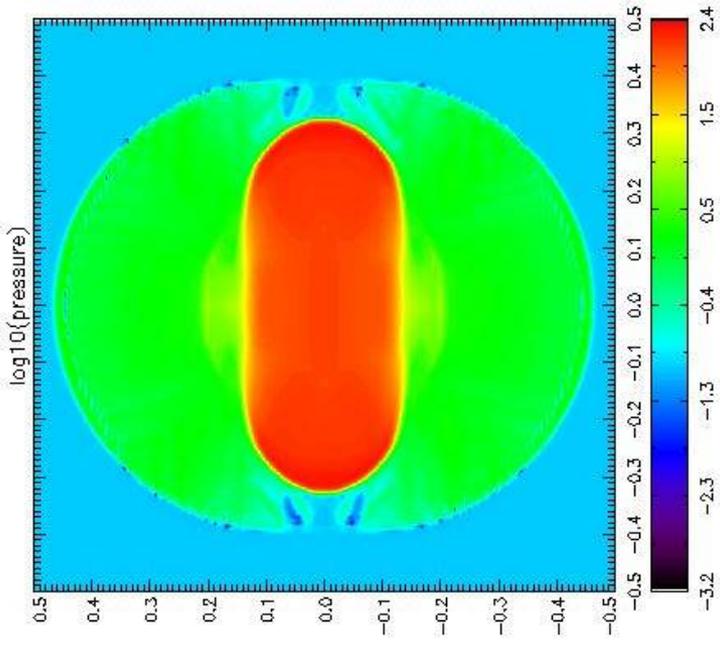
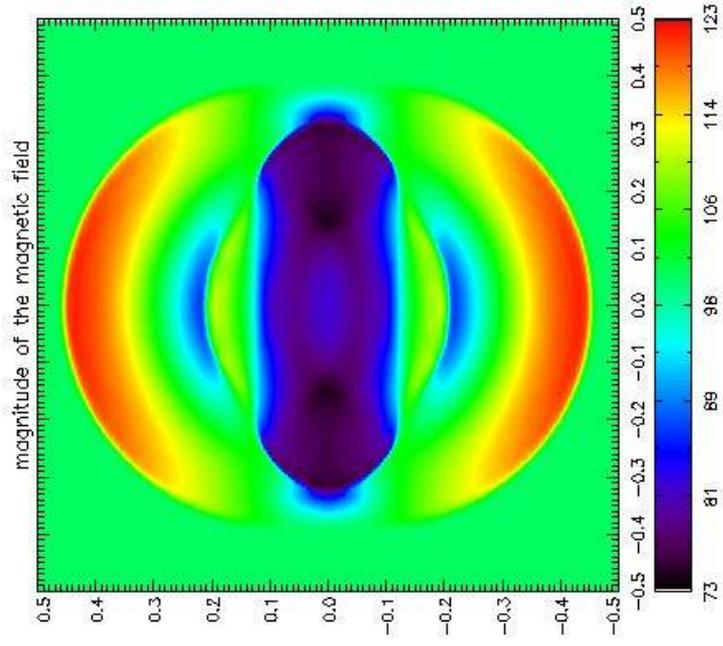
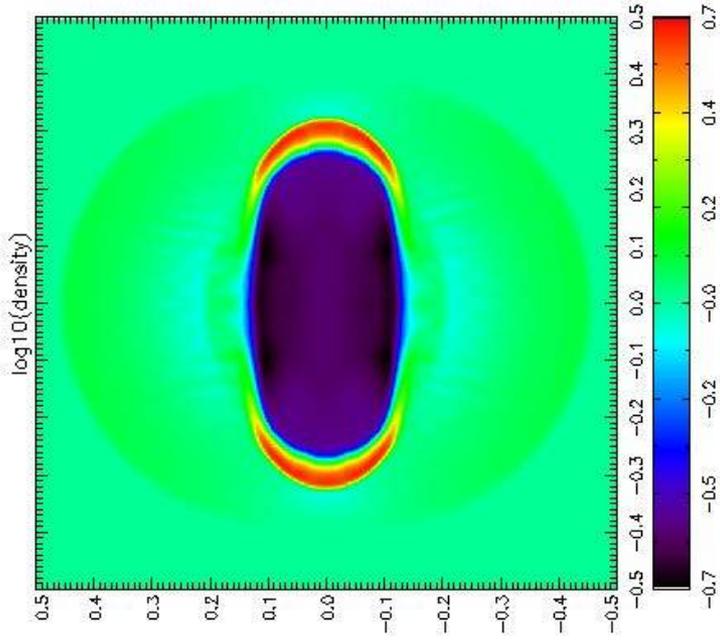
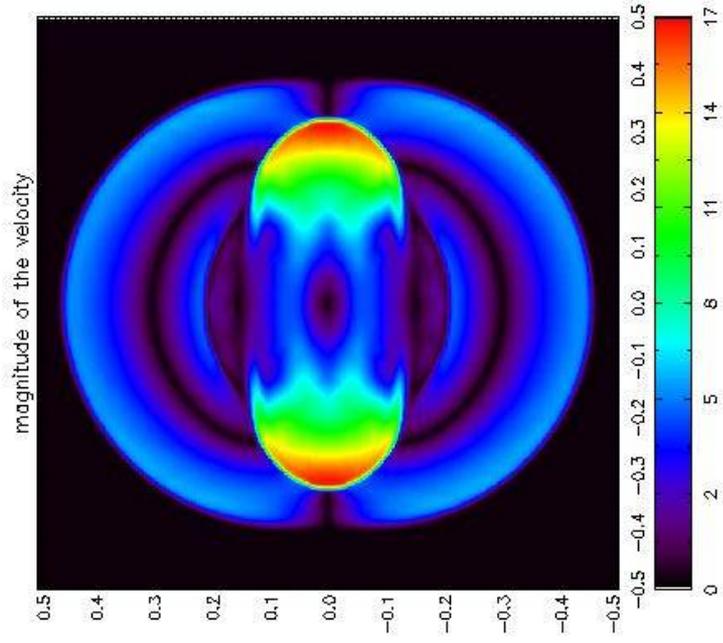

Fig. 3) Blast problem in 2d where the top two panels show the $\log_{10}$ of the density and pressure and the bottom two panels show the magnitudes of the velocity and the magnetic field. The fourth order WENO reconstruction with third order accurate RK time-stepping along with an HLLE Riemann solver was used.